\documentclass[superscriptaddress,groupedaddress,nofootnoteinbib,12pt]{article}
\usepackage{color}
\usepackage{graphicx}
\usepackage{dcolumn}
\usepackage{bm}     
\usepackage{bbm}  
\usepackage{amssymb}
\usepackage{amsmath}
\usepackage{sectsty}
\usepackage{colortbl}
\usepackage{latexsym}
\usepackage{float}
\usepackage{ifthen}
\usepackage{caption,subfig}
\usepackage{enumerate}
\usepackage{url}
\usepackage{caption,subfig}
\usepackage{feynmf}	
\usepackage{jcappub}

\def\be{\begin{equation}}
\def\ee{\end{equation}}
\def\ba{\begin{eqnarray}}
\def\ea{\end{eqnarray}}
\def\beq{\begin{eqnarray}}
\def\eeq{\end{eqnarray}}

\def\mpl{M_{\rm Pl}}

\def\E{\mathcal{E}}

\def\d{\mathrm{d}}
\def\p{{\cal P}}

\def\K{{\cal K}}

\def\L*{{\cal L}_*}
\def\L{\mathcal{L}}
\def\({\left(}
\def\){\right)}

\def\p{\partial}
\def\mn{_{\mu \nu}}
\def\ab{_{ab}}
\def\stu{St\"uckelberg }
\def\p{\partial}

\def\<{\langle}
\def\>{\rangle}

\def\cs2{c_{s}^{2}}

 \def\p{\partial}

\def\be{\begin{equation}}
\def\ee{\end{equation}}
\def\ba{\begin{eqnarray}}
\def\ea{\end{eqnarray}}
\def\beq{\begin{eqnarray}}
\def\eeq{\end{eqnarray}}

\def\mpl{M_{\rm Pl}}

\def\E{\mathcal{E}}

\def\d{\mathrm{d}}
\def\p{{\cal P}}

\def\K{{\cal K}}

\def\L*{{\cal L}_*}
\def\L{\mathcal{L}}
\def\({\left(}
\def\){\right)}

\def\p{\partial}
\def\mn{_{\mu \nu}}
\def\stu{St\"uckelberg }
\def\p{\partial}

\def\<{\langle}
\def\>{\rangle}

 \def\be   {\begin{equation}}   \def\ee   {\end{equation}}

 \def\ba  {\begin{eqnarray}}   \def\ea  {\end{eqnarray}}

\renewcommand{\thesubsection}{\arabic{section}.\arabic{subsection}}

\begin{document}
\hspace{5.2in} \mbox{NORDITA-2014-118}\\\vspace{-1.03cm}

\title{Quantum corrections in massive bigravity and new effective composite metrics}

\author{Lavinia Heisenberg$^{a,b}$}
\affiliation{$^{a}$Nordita, KTH Royal Institute of Technology and Stockholm University, \\
Roslagstullsbacken 23, 10691 Stockholm, Sweden}
\affiliation{$^{b}$Department of Physics \& The Oskar Klein Centre, \\
AlbaNova University Centre, 10691 Stockholm, Sweden}

	\emailAdd{laviniah@kth.se}

\abstract{We compute the one-loop quantum corrections to the interactions between the two metrics of the ghost-free massive bigravity. When considering gravitons running in the loops, we show how the structure of the interactions gets destabilized at the quantum level, exactly in the same way as in its massive gravity limit. A priori one might have expected a better quantum behavior, however the broken diffeomorphism invariance out of the two initial diffeomorphisms in bigravity has similar consequences at the quantum level as the broken diffeomorphism in massive gravity. From lessons of the generated quantum corrections through matter loops we propose yet other types of effective composite metrics to which the matter fields can couple. Among these new effective metrics there might be one or more that could provide interesting phenomenology and important cosmological implications.
}

\maketitle

\section{Introduction}

Independent cosmological observations such as supernovae, CMB, baryon acoustic oscillations and lensing indicate an accelerated expansion of the Universe, driven by something that we call "Dark Energy". The given name reflects the fact that its origin is still unknown despite a significant theoretical and observational efforts. The expansion could be for instance due to a small cosmological constant $\lambda$ with a constant energy density. However, under the assumption that the cosmological constant is generated by the vacuum energy density, we can use the standard quantum field theory techniques to compute the vacuum energy density caused by fluctuating quantum fields. The result is puzzling. It differs from the observational bounds by $~120$ orders of magnitude. This large discrepancy between the theoretically computed high energy density of the vacuum and the observational value constitutes the cosmological constant problem \cite{Weinberg:1988cp}. The accelerated expansion of the Universe could also be due to  new dynamical degrees of freedom, either by invoking new fluids with negative pressure or by changing the geometrical part of Einstein's equations. Among the latter class of approaches, the infrared-modifications of gravity offer promising and exciting new ways for not only addressing the late-time acceleration enigma but also tackling the cosmological constant problem. Important representatives of this type of infrared-modifications are massive gravity and higher-dimensional setups. In the context of higher dimensional theories the Dvali-Gabadadze-Porrati (DGP) \cite{Dvali:2000hr} model has revolutionized the early stages of large scale modified theories of gravity. For about ten years later the community working on infrared-modifications has witnessed another revolutionary result by the work of de Rham-Gabadadze-Tolley (dRGT) \cite{deRham:2010kj}, who successfully extended the mass term in massive gravity to the non-linear level without invoking ghostly degrees of freedom, which was a challenge over forty years (for extended reviews see \cite{deRham:2014zqa, Hinterbichler:2011tt}). \\

Parallel to these new exciting achievements, Galilean invariant interactions were proposed to extend the decoupling limit of DGP-gravity \cite{Nicolis:2008in}. The helicity-0 mode $\pi$ of the DGP model has the invariance under internal galileon- and shift transformations $\pi\to\pi+b_\mu x^\mu+c$. Together with the postulate of the absence of ghosts these symmetries restrict the allowed effective Galileon Lagrangian (for an extensive review see Ref.~\cite{deRham:2012az}). A crucial property of the Galileon is the non-renormalization theorem which ensures that the Galileon coupling constants are technically natural and stable under quantum corrections \cite{Luty:2003vm,Nicolis:2004qq,Hinterbichler:2010xn,deRham:2012ew,Heisenberg:2014raa}. In the context of massive gravity, the Galileon-type interactions naturally arise in the decoupling limit and provides rich phenomenology \cite{deRham:2010ik,deRham:2010tw, Berezhiani:2013dca, Berezhiani:2013dw,deFromont:2013iwa,Burrage:2011cr}. The nice properties of the Galileon can also be generalized to higher spin fields, like vector fields..etc \cite{Deffayet:2013tca,Deffayet:2010zh, Heisenberg:2014rta, Tasinato:2014eka,Jimenez:2013qsa,Jimenez:2014rna}. There has been successful attempts in generalizing the Galileon to the covariant Galileon on non-flat backgrounds. First covariantization consisted on the explicit second order equations of motion sacrificing the Galileon symmetry \cite{Deffayet:2009mn, Deffayet:2009wt, deRham:2010eu} (even though generalizations to the maximally symmetric backgrounds did still share a generalized Galileon symmetry \cite{Burrage:2011bt, Goon:2011qf}). If one is willing to give up on the restriction of second order equations of motion (but still avoiding ghost instabilities) one can construct covariant Galileon interactions with promising new phenomenology \cite{Gleyzes:2014dya,Gleyzes:2014qga, Gao:2014soa,Fasiello:2014aqa}. Interestingly, a subclass of covariant Galileon interactions naturally arise from the covariantization of the decoupling limit of massive gravity \cite{deRham:2011by, Heisenberg:2014kea}. \\

The potential interactions in the dRGT theory was constructed in a way such that the Boulware-Deser (BD) ghost remains absent at the non-linear level. This is guarantied by the presence of a fundamental matrix constructed out of the squared root of $\hat{g}^{-1}\hat{f}$, where $\hat{g}$ represents the dynamical metric and $\hat{f}$ the reference metric \cite{deRham:2010kj}. Of course it is a natural question whether or not this very specific structure of the potential is stable under quantum corrections. These questions have been explored in  \cite{Park:2010rp,Buchbinder:2012wb,deRham:2013qqa, deRham:2014naa}. 
Following the aforementioned motivations, there has been numerous investigations of the dRGT massive gravity concerning the late-time accelerated expansion of the Universe and other  phenomenological aspects  \cite{deRham:2010tw,PhysRevD.84.124046,PhysRevLett.109.171101,deRham:2011by,Chamseddine:2011bu,Koyama:2011xz,Koyama:2011wx,Gumrukcuoglu:2011zh,Gratia:2012wt,Vakili:2012tm,Kobayashi:2012fz,Fasiello:2012rw,Volkov:2012zb,Tasinato:2012ze,DeFelice:2013bxa,Fasiello:2013woa,Heisenberg:2014kea,Comelli:2013tja,Motloch:2014nwa,Berezhiani:2011mt}. Even if dRGT theory is a IR modification of GR, the lessons learned there can also be applied to UV modifications of GR \cite{Jimenez:2014fla}.
Furthermore, the dRGT theory was extended to its bimetric version by promoting the reference metric $\hat{f}$ to a dynamical metric through an additional kinetic term for $\hat{f}$ \cite{Hassan:2011zd}.
Independently of the dynamics of the reference metric, it is a mandatory question of how the two metrics can be coupled to the matter sector in a consistent way, meaning without invoking the BD ghost.
This question has been already explored in a multitude of very interesting works in \cite{Khosravi:2011zi, Akrami:2012vf,Akrami:2013ffa,Tamanini:2013xia,Akrami:2014lja, deRham:2014naa,Yamashita:2014fga,Noller:2014sta,Hassan:2014gta,Schmidt-May:2014xla,Enander:2014xga,Gumrukcuoglu:2014xba,Mukohyama:2014rca,Soloviev:2014eea}. The ghost-freedom must be maintained at the quantum level as well, at least below the cut-off scale of the theory. Thus, the quantum behavior will deliver additional constraints on the coupling to matter. One natural way of coupling the matter field is to couple the matter sector to only one metric, and not to both metrics simultaneously. In this case the classical ghost-freedom remains also at the quantum level as shown in \cite{deRham:2014naa}. This is due to the fact, that the quantum corrections contribute only in form of two cosmological constants for the two metrics. Even if the matter sector couples to only one of the two metrics, quantum corrections will generate a coupling to the other metric and it will be important to investigate at which scale this new coupling will be generated. This will be one of the questions that we will ask. Moreover, even if it is tempting to couple the matter field to both metrics simultaneously, one immediately faces the appearance of the BD ghost already at the classical level. On top of that the quantum corrections detune the specific potential structure at an arbitrarily low scale. Hence, this way of coupling would render the theory sick. Under the requirement that the ghost-free potential structure is not detuned by the quantum corrections, one can construct a new composite effective metric built out of both metrics, through which the matter field can couple  \cite{deRham:2014naa}. This coupling does not introduce the ghost degree of freedom at least up to the strong coupling scale and can be used as a perfectly valid effective field theory with a cut-off above the strong coupling scale. Following the philosophy of \cite{deRham:2014naa} we will use the lessons learned about the quantum corrections coming from matter loops in order to introduce yet other types of effective metrics through which the matter field can couple to both metrics at the same time. Moreover, we will also consider the quantum corrections generated by purely graviton loops and show the detuning of the specific potential interactions in parallel to what happens in massive gravity \cite{deRham:2013qqa}. \\

\section{Ghost-free Bigravity}
\label{sec:dRGT}
In this section we will first review the ghost-free interactions in the theory of massive bigravity and setup the framework in which we will perform the one-loop computation. Our starting point is the action for bimetric gravity and the matter action sourcing for gravity  \cite{deRham:2010kj,Hassan:2011vm}

\begin{equation}\label{action_bigravity}
\mathcal{S}_{\rm BG} = \int \mathrm{d}^4x \big[- \frac{M_p^2}{2} \sqrt{g}\left(R[g]+\frac{m^2}{2}\sum_n \alpha_n{\cal U}[\cal K]\right) -  \frac{M_{\rm f}^2}{2} \sqrt{f}R_f +\mathcal L_m(g,f,\psi_m) \big]
\end{equation}
where the potential interactions are given by \cite{deRham:2010ik,deRham:2010kj}
\begin{eqnarray}
\mathcal{U}_0[\mathcal{K}] &=&\frac{1}{24} \mathcal{E}^{\mu\nu\rho\sigma}  \mathcal{E}_{\mu\nu\rho\sigma} =1\nonumber\\
\mathcal{U}_1[\mathcal{K}] &=&\frac{1}{6} \mathcal{E}^{\mu\nu\rho\sigma}  \mathcal{E}^{\alpha}_{\;\;\;\nu\rho\sigma} \K_{\mu\alpha} =[\K] \nonumber\\
\mathcal{U}_2[\mathcal{K}] &=&\frac{1}{4} \mathcal{E}^{\mu\nu\rho\sigma}  \mathcal{E}^{\alpha\beta}_{\;\;\;\;\; \rho\sigma} \K_{\mu\alpha} \K_{\nu\beta} = \frac{1}{2}\left( [\K]^2-[\K^2]\right), \nonumber\\
\mathcal{U}_3[\mathcal{K}] &=&\frac{1}{6}  \mathcal{E}^{\mu\nu\rho\sigma}  \mathcal{E}^{\alpha\beta\kappa}_{\;\;\;\;\;\;\; \sigma} \K_{\mu\alpha} \K_{\nu\beta}  \K_{\rho\kappa}=\frac{1}{6}\left([\K]^3-3[\K][\K^2]+2[\K^3]\right),  \nonumber\\
\mathcal{U}_3[\mathcal{K}] &=&\frac{1}{24}  \mathcal{E}^{\mu\nu\rho\sigma}  \mathcal{E}^{\alpha\beta\kappa\gamma} \K_{\mu\alpha} \K_{\nu\beta}  \K_{\rho\kappa}  \K_{\sigma\gamma} =\frac{1}{24}\left([\K]^4-6[\K]^2[\K^2]+3[\K^2]^2+8[\K][\K^3]-6[\K^4]\right), \nonumber
\end{eqnarray}
where $\mathcal{E}$ stands for the Levi-Cevita tensor. The tensor $\K$ has a very non-trivial structure in form of a square root
\begin{equation}
\K^\mu _\nu[g,f] =\delta^\mu_\nu - \left(\sqrt{g^{-1}f}\right)^\mu_\nu \,.
\end{equation}
In comparison to massive gravity, the potential term here represents the potential for both metrics and $\hat{f}$ is a dynamical metric as well. The ghost absence has been also successfully proven for the case of bigravity \cite{Hassan:2011zd}.\\

The same square-root structure which guaranties the ghost absence, makes life very hard. This mathematically cumbersome structure can be avoided using the vielbein language \cite{Nibbelink:2006sz,Hinterbichler:2012cn,Deffayet:2012zc, Gabadadze:2013ria,Ondo:2013wka} since the vielbein is like the 'square-root' of the metric. The ghost-free potential becomes a simple polynomial in the vielbein formalism and contains interactions up to quartic order in the vielbein fields. Therefore, we will work in a symmetric-vielbein inspired language in the euclidean space in a similar way as was done in \cite{deRham:2013qqa}. Thus, the metrics are expressed as
\begin{eqnarray}
\label{vielbeinFormgf}
g_{ab}=\left(\bar{\Gamma}_{ab}+\frac{h_{ab}}{\mpl}\right)^2
\equiv
\left(\bar \Gamma_{ac}+\frac{h_{ac}}{\mpl}\right)\left(\bar \Gamma_{db}+\frac{h_{db}}{\mpl}\right)\delta^{cd} \nonumber\\
f_{ab}=\left(\bar{ \mathcal{Q}}_{ab}+\frac{l_{ab}}{M_{\rm f}}\right)^2
\equiv
\left(\bar{\mathcal{Q}}_{ac}+\frac{l_{ac}}{M_{\rm f}}\right)\left(\bar{\mathcal{Q}}_{db}+\frac{l_{db}}{M_{\rm f}}\right)\delta^{cd} 
\end{eqnarray}
with $\bar g_{ab}=\bar{\Gamma}\ab^2=\bar{\Gamma}_{ac}\bar{\Gamma}_{bd}\delta^{cd}$ and $\bar f_{ab}=\bar{\mathcal{Q}}\ab^2=\bar{\mathcal{Q}}_{ac}\bar{\mathcal{Q}}_{bd}\delta^{cd}$ being the background metrics for $g_{ab}$ and $f_{ab}$ respectively and the fluctuations are denoted by $h\ab$ and $l\ab$. Of course the background metrics $\bar g_{ab}$ and $\bar f_{ab}$ do not need to be flat, however, for simplicity for most of the computations we will assume flat backgrounds. When working around flat background metrics $\bar{\Gamma}_{ab}^2=\delta_{ab}$ and $\bar{\mathcal{Q}}_{ab}^2=\delta_{ab}$, following expressions will be useful throughout the paper 
\ba
g_{ab}&=&\delta_{ab}+\frac{2}{\mpl}h_{ab}+\frac{1}{\mpl^2}h_{ac}h_{bd}\delta^{cd}\nonumber\\
f_{ab}&=&\delta_{ab}+\frac{2}{M_{\rm f}}l_{ab}+\frac{1}{M_{\rm f}^2}l_{ac}l_{bd}\delta^{cd} \nonumber\\
g^{ab}&=&\delta^{ab}-\frac{2}{\mpl}h^{ab}+\frac{3}{\mpl^2}h^a_c h^{cb}+\cdots \nonumber\\
f^{ab}&=&\delta^{ab}-\frac{2}{M_{\rm f}}l^{ab}+\frac{3}{M_{\rm f}^2}l^a_c l^{cb}+\cdots \nonumber\\
\label{eq:defmetricsfluctuations}
\ea
With this form of the fluctuations the squared root of the determinants then become
\ba\label{sqrtgandf}
\sqrt{g}&=&1+\frac{[h]}{\mpl}+\frac1{2\mpl^2}([h]^2-[h^2])+\cdots \nonumber\\
\sqrt{f}&=&1+\frac{[f]}{M_{\rm f}}+\frac1{2M_{\rm f}^2}([f]^2-[f^2])+\cdots 
\ea
We will perform the computation of the one-loop quantum corrections using dimensional regularization. Therefore, the contributions in form of a measure term in the path integral arising from the field redefinitions $g_{ab}$ to $h_{ab}$ and $f_{ab}$ to $l_{ab}$ will not be considered here since they generate power law divergent quantum corrections. Physics is insensitive to field redefinitions, therefore they will be governed by the logarithmic runnings (see \cite{deRham:2014wfa} for an interesting discussion on this).

\subsection{Quantum corrections in the decoupling limit}
Let us first have a look at the quantum corrections to the potential arising in the decoupling limit. This is the first thing to be checked. If the quantum corrections already destabilize the decoupling limit itself, then there is no need to look at the full theory and the theory would be rendered sick. In massive gravity, the non-renormalization theorem protects the interactions within the decoupling limit. If the same is true for the bigravity case, then the theory would be safe under quantum corrections at least within the decoupling limit. Once this has been sorted out, the full theory at the quantum level can be studied as the next step. The decoupling limit provides a framework in which the most important physical properties of the theory are visible since the individual degrees of freedom decouple from each other. 
In bigravity the interaction between the two metrics $g_{ab}$ and $f_{ab}$ breaks the two copies of diffeomorphisms down to one, such that in the decoupling limit the interactions are governed by decoupled two helicity-2 modes $h_{ab}$,  two helicity-2 modes $l_{ab}$, two helicity-1 modes $A_a$ and one helicity-0 mode $\pi$ accounting in total seven propagating helicity modes. 
As it is already visible in the action \ref{action_bigravity}, the two metrics come in at their own Planck masses $\mpl$ and $M_{\rm f}$, therefore the decoupling limit of bigravity represents the limit in which (see \cite{Fasiello:2013woa} for the first derivation of the decoupling limit in bigravity)
\begin{eqnarray}
\mpl\to\infty,\;\; M_{\rm f}\to\infty,\;\; m\to0 \;\;\;\;\;\; {\rm and} \;\;\;\;\;\;  \frac{\mpl}{M_{\rm f}}={\rm const}
\end{eqnarray}
The resulting theory in this limit contains interactions at the lowest energy scale between the two helicity-2 fields $h_{ab}$ and $l_{ab}$ and the helicity-0 scalar field $\pi$ in the following form (please see \cite{Fasiello:2013woa} for a detailed derivation)
\begin{eqnarray}
S&=&\int \d ^4 x \left[h^{ab}\hat{\mathcal{E}}^{abcd}h_{cd}+l^{ab}\hat{\mathcal{E}}^{abcd}l_{cd}-\Lambda_3^3\sum_{n=0}^{4}h^{ab}X^{(n)}_{ab}-\frac{M_p}{M_f}\Lambda_3^3\sum_{n=0}^{4}l^{ab}\tilde Y^{(n)}_{ab}\right]
\end{eqnarray}
where $\hat{\mathcal{E}}$ is the Lichnerowicz operator
\begin{eqnarray}
\hat{\mathcal{E}}^{cd}_{ab} h_{cd}=-\frac 12 \(\Box h_{ab}-2\p_c \p_{(a}h^c_{b)}+\p_a\p_b h-\delta_{ab} (\Box h-\p_c\p_d h^{cd})\)\,,
\label{eq:epsilondef}
\end{eqnarray}
and the $X_{ab}$ and $\tilde Y_{ab}$ encode the derivative interactions of order $n$ in the helicity-0 field $\pi$
\begin{eqnarray}
X^{(n)}_{ab}&=&-\frac12 \frac{\hat\beta_n}{(3-n)!n!}\mathcal E_a^{\;\;\cdots}\mathcal E_b^{\;\;\cdots}(\delta+\Pi)^n\delta^{3-n} \nonumber\\
\tilde Y^{(n)}_{ab}&=&-\frac12 \frac{\hat\beta_n}{(4-n)!(n-1)!}\mathcal E_a^{\;\;\cdots}\mathcal E_b^{\;\;\cdots}\delta^{(n-1)}(\delta+\Sigma)^n\delta^{4-n} 
\end{eqnarray}
where $\hat \beta_n=M_p^2\beta_n$. The authors in  \cite{Fasiello:2013woa} use the $\beta_n$ notation which we borrow here for the sake of this section (the relation between the parameters $\hat \beta_n$ and $\alpha_n$ are given in equation 2.14 of  \cite{Fasiello:2013woa}). Furthermore, $\Pi$ and $\Sigma$ stands for $\Pi_{ab}=\partial_a\partial_b\pi/\Lambda_3^3$ and $\Sigma_{ab}=\partial_a\partial_b\rho/\Lambda_3^3$ respectively and $\rho$ is the dual description of $\pi$ via field redefinitions related in a form
\begin{eqnarray}
(\delta+\Sigma)=(\delta+\Pi)^{-1}
\end{eqnarray}
The interactions between the helicity-2 field $h_{ab}$ and the helicity-0 field $h^{ab}X^{(n)}_{ab}$ are exactly the same as in the decoupling limit of massive gravity. In \cite{deRham:2012ew} it has been shown that these interactions are protected from quantum corrections via the non-renormalization theorem. This property is thanks to the antisymmetric structure of the interactions. In the decoupling limit of bigravity we have the additional interactions between the helicity-2 field $l\mn$ and the helicity-0 field via $l^{ab}\tilde{Y}^{(n)}_{ab}$. However, it is easy to convince ourselves that exactly the same argumentation for the non-renormalization theorem used in massive gravity applies also here in bigravity. The essential operator for the non-renormalization theorem is the Levi-Civita tensor which is also contained in the interactions $l^{ab}\tilde{Y}^{(n)}_{ab}$. Exactly this property will guarantie that any external particle contracted with any field with or without derivatives in a vertex contributes to a two-derivatives operator acting on this external
particle, which gives rise to counter terms that do not have the same structure as the classical interactions and hence do not renormalize. In bigravity we basically have two copies of the same non-renormalization theorem, namely for $h^{ab}X^{(n)}_{ab}$ and $l^{ab}\tilde Y^{(n)}_{ab}$ interactions.
Take for instance the interaction $ l^{ab}\tilde Y^{(n)}_{ab} \supseteq l^{ab} \mathcal{E}_{a}^{\;\;cek} \mathcal{E}_{b\;\;\;\;k}^{\;\;df}(\delta_{cd}+\Sigma_{cd})(\delta_{ef}+\Sigma_{ef}$). The part with the $\delta$'s correspond to a tadpole contribution and a kinetic term for $\rho$ such that the only non-trivial interaction will come from $ l^{ab} \mathcal{E}_{a}^{\;\;cek} \mathcal{E}_{b\;\;\;\;k}^{\;\;df}\Sigma_{cd}\Sigma_{ef}$. Now contract an external helicity-2 particle $l^{ab}$ with momentum  $q_a$ with the helicity-2 field coming without derivatives in this interaction at a vertex while letting the other two $\rho$-particles dual to the helicity-0 field $\pi$ run in the loop with momenta $p_a$ and $(q+p)_a$. The contribution of this vertex gives
\begin{eqnarray}
\mathcal A
 \propto  \int \frac{\mathrm{d}^4 k}{(2\pi)^4} G_p \,
 G_{p+q}\  \epsilon_{ab}\, \mathcal{E}^{ace k} \,
\mathcal{E}_{\;\;\;\;k}^{bdf}
\, \, p_c \, p_d \, (q+p)_e \, (q+p)_f \cdots \,,
\end{eqnarray}
where $\epsilon_{ab}$ stands for the spin-2 polarization tensor and $G_p= p^{-2}$ for the Feynman massless propagator of the $\rho$ field. Exactly in the same way as it happens in the decoupling limit of massive gravity, the Levi-Civita antisymmetric structure of the vertex enforces that only the terms with at least two powers of the external helicity-2 momentum $q_e q_f$ contributes to the scattering amplitude \cite{deRham:2012ew}. The same is true for the remaining interactions between the helicity-0 field and the helicity-1 field in the decoupling limit of massive gravity (their exact form is given in  \cite{Fasiello:2013woa}, however only their Levi-Civita antisymmetric structure matters for the non-renormalization theorem). Thus, the decoupling limit of bigravity is protected from quantum corrections. This is a trivial generalization of the non-renormalization theorem to the case of bigravity.


\subsection{Propagators for the massless and massive modes in the unitary gauge}

Now that we have established the non-renormalization argument in the decoupling limit above, we can investigate the quantum corrections in the full non-linear theory. We will perform the analysis in the 
unitary gauge with vanishing \stu fields, i.e. $\Phi^a=x^\alpha\delta^a_{\ \alpha}$. In contrast to massive gravity, we have two spin-2 fields. In order to compute the one-loop quantum corrections, we have to specify the mass spectrum of bigravity. For that we will split the mass spectrum of biravity into massive and massless spin-$2$ fluctuations around flat euclidean backgrounds $\bar g_{ab}=\delta_{ab}=\bar f_{ab}$. To be precise, for the mass spectrum we will perform the metric perturbations in \ref{eq:defmetricsfluctuations}. Then, the fundamental matrix of the theory in terms of the perturbations is given by
\begin{eqnarray}\label{Kuv_pert1}
\mathcal{K}^a_{\;b}=-\frac{l^a_{\;b}}{M_{\rm f}}+\frac{h^a_{\;b}}{\mpl}-\frac14\frac{l^a_{\;c}l^c_{\;b}}{M_{\rm f}^2}-\frac54\frac{h^a_{\;c}h^c_{\;b}}{\mpl^2}+\frac32\frac{h^a_{\;c}l^c_{\;b}}{\mpl M_{\rm f}} + \cdots
\end{eqnarray}
The action for the bigravity \ref{action_bigravity} up to quadratic order in the perturbations becomes
\begin{eqnarray}
S&=&\int \mathrm{d} ^4 x \left\{h_{ab}\hat{\mathcal{E}}^{abcd}h_{ab}+l_{ab}\hat{\mathcal{E}}^{abcd}l_{cd} \right. \nonumber\\
&& +\frac{\alpha_2m^2}{4}\left[ ([h^2]-[h]^2)+\frac{\mpl^2}{M_{\rm f}^2}([l^2]-[l]^2)-2\frac{\mpl}{M_{\rm f}}(h^{ab}l_{ab}-[h][l]) \right] \left. \right\}
\end{eqnarray}
We can now diagonalize these interactions by making the following change of variables 
\begin{eqnarray}\label{fluctuations_massmodes}
h_{ab}&\to& \mpl(w_{ab}+v_{ab}) \nonumber\\
l_{ab}&\to& M_{\rm f} (w_{ab}-v_{ab})
\end{eqnarray}
such that the action at quadratic order in perturbations becomes \cite{Hassan:2011zd}  
\begin{equation}
S=\int \mathrm{d} ^4 x \left\{w_{ab}\hat{\mathcal{E}}^{abcd}w_{cd}+v_{ab}\hat{\mathcal{E}}^{abcd}v_{cd}+\alpha_2m^2\mpl^2\left[[v^2]-[v]^2\right] \right\}\,.
\end{equation}
In the unitary gauge $v_{ab}$ encodes all the five physical degrees of freedom of a massive spin-2 fluctuation (the two helicity-2, the two helicity-1 and the helicity-$0$ modes), and $w_{ab}$ encodes the two helicity-2 modes of the massless fluctuation.
The Feynman propagator for the massless spin-2 fluctuation $w_{ab}$ is given by
\begin{eqnarray}
\label{GR_propagator}
G^{ (w)}_{abcd}=\langle w_{ab}(x_1) w_{cd}(x_2)\rangle=f^{(w)}_{abcd}  \,
\int \frac{\mathrm{d}^4 k}{(2\pi)^4}
\frac{e^{i k \cdot  \left(x_1- x_2 \right)}}{k^2},
\end{eqnarray}
where the polarization structure has the usual prefactor of $1/2$
\begin{eqnarray}
\label{f}
f^{(w)}_{abcd}=\delta_{a(c}\delta_{bd)}-\frac 12  \delta_{ab} \delta_{cd}\,.
\end{eqnarray}
with $\delta_{a(c}\delta_{bd)}\equiv\frac{1}{2} \delta_{ac} \delta_{bd}+\frac{1}{2}\delta_{ad}\delta_{bc}.$ The massive spin-2 field,  on the other hand, has the Feynman propagator
\begin{eqnarray}
\label{MG_propagator}
G^{(v)}_{abcd}&=&\langle v_{ab}(x_1) v_{cd}(x_2)\rangle=f^{(v)}_{abcd} \int \frac{\mathrm{d}^4 k}{(2\pi)^4}\frac{e^{i k \cdot \left(x_1 - x_2\right)}}{k^2+m^2}\,,
\end{eqnarray}
with the prefactor of $1/3$ in the polarization structure
\begin{eqnarray}
f^{(v)}_{abcd}=\left(\tilde \delta_{a(c}\tilde \delta_{bd)}-\frac 13 \tilde \delta_{ab} \tilde  \delta_{cd}\right)
\hspace{15pt}{\rm where}\hspace{15pt}
\tilde \delta_{ab}=\delta_{ab}+\frac{k_a k_b}{m^2}\,.
\label{eq:fmassive}
\end{eqnarray}


\section{Graviton loops}
\label{sec:GravitonLoops}

In this section we will study the quantum corrections generated by the graviton loops. We will be concentrating on one-loop diagrams. We will be only interested in the IR limit of the theory, therefore consider zero external momenta. We will apply dimensional regularization and thus focus only on the running of the interaction couplings.

\subsection{Preparative study}

The quantum corrections in the decoupling limit of bigravity follow the same non-renormalization theorem as in massive gravity. In the decoupling limit the coupling to matter fields are suppressed as $M_f\to\infty$ and $M_p\to\infty$. So the decoupling limit of the bimetric theory is completely safe. Now here we want to investigate the quantum corrections of the full non-linear bimetric theory coming from purely graviton loops. \\

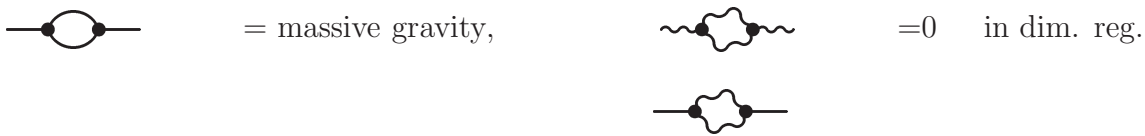
\begin{figure}[!htb]\vspace{40pt}
\begin{center}
\begin{fmffile}{seperatefg}
\parbox{30mm}{\begin{fmfgraph*}(50,30)
	             \fmfleft{i1}
	            \fmfright{o1}
                \fmf{plain}{i1,v1}
                \fmf{plain}{v2,o1}
                \fmfdot{v1}
                \fmfdot{v2}
                \fmf{plain,left=0.7,tension=0.4}{v1,v2,v1}
\end{fmfgraph*}} = {\rm massive gravity}, \hspace{55pt} 
\parbox{30mm}{\begin{fmfgraph*}(50,30)
	            \fmfleft{i1}
	            \fmfright{o1}
                \fmf{wiggly}{i1,v1}
                \fmf{wiggly}{v2,o1}
                \fmfdot{v1}
                \fmfdot{v2}
                \fmf{wiggly,left=0.7,tension=0.4}{v1,v2,v1}
\end{fmfgraph*}} =0 \;\;\; {\rm in dim. reg.}\\
 \hspace{145pt} \parbox{30mm}{\begin{fmfgraph*}(50,30)
	            \fmfleft{i1}
	            \fmfright{o1}
                \fmf{plain}{i1,v1}
                \fmf{plain}{v2,o1}
                \fmfdot{v1}
                \fmfdot{v2}
                \fmf{wiggly,left=0.7,tension=0.4}{v1,v2,v1}
\end{fmfgraph*}} 
\end{fmffile}
\caption{One-loop quantum contributions from gravitons: the straight line denotes the massive mode and the curly line the massless mode. Loops in which only the massless mode runs give zero contributions.}
\end{center}
\label{seperatefg}
\end{figure}
Our starting point will be expanding the potential interactions \ref{action_bigravity} in terms of the fluctuations \ref{eq:defmetricsfluctuations}. Before starting the computation we can already gain a lot by noting that the separate $h_{ab}$ and $l_{ab}$ interactions without any mixing between them will give rise to the same results as we obtained in massive gravity. The crucial point for that is that once we express the fluctuations in terms of the mass eigenstates $v_{ab}$ and $w_{ab}$ then the one loop contributions in which only the massless degree of freedom run, will give rise to zero contributions in dimensional regularization, i.e. in the cut-off regularization there is no logarithmic divergences. Thus, if we have graviton one-loop with only the massless mode $w_{ab}$ running in it, then this will give zero contribution since we have a contribution of the form
\begin{equation}
\int d^d k \frac{A(k,\mu)}{k^2}=0 \;\;\;\;\;\;\;\  \text{ in dimensional regularization}
\end{equation}
If we consider one loop diagrams in which either only the fluctuations of the metric $g_{ab}$ or only fluctuations of the metric $f_{ab}$ come in, then the quantum corrections from the massless mode $w_{ab}$ will give zero contribution while the one for the massive mode $v_{ab}$ will end up giving the same contribution as in massive gravity. \\
Therefore, we already gained a lot by realizing this and we only need to concentrate on the contributions coming from the mixed diagrams. But from the mixed diagrams we only need to consider those cases in which only the massive mode runs or where massive and massless mode run in the same loop but never the mixed diagrams with purely massless mode running in the loop.
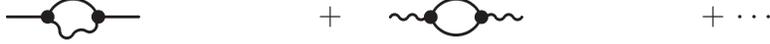
\begin{figure}[!htb]\vspace{40pt}
\begin{center}
\begin{fmffile}{mixedfg}
\parbox{40mm}{\begin{fmfgraph*}(50,30)
	             \fmfleft{i1}
	            \fmfright{o1}
                \fmf{plain}{i1,v1}
                \fmf{plain}{v2,o1}
                \fmfdot{v1}
                \fmfdot{v2}
                \fmf{plain,left=0.7,tension=0.4}{v1,v2}
                 \fmf{wiggly,left=0.7,tension=0.4}{v2,v1}
\end{fmfgraph*}} + \hspace{10pt}
\parbox{40mm}{\begin{fmfgraph*}(50,30)
	            \fmfleft{i1}
	            \fmfright{o1}
                \fmf{wiggly}{i1,v1}
                \fmf{wiggly}{v2,o1}
                \fmfdot{v1}
                \fmfdot{v2}
                \fmf{plain,left=0.7,tension=0.4}{v1,v2,v1}
\end{fmfgraph*}} + $\cdots$
\end{fmffile}
\end{center}
\caption{Only mixed diagrams in which $w$ and $v$ run, will give new non-trivial contributions.}
\label{mixedfg}
\end{figure}
We also expect here that the separate diagrams will give rise to detuning of the potential interactions as in massive gravity. Since the $f_{ab}$ is dynamical, we have one full diffeomorphism invariance which might give rise to a better behaviour at the quantum level and some cancellations might be possible. However, we will see that this is not the case, at least among the diagrams constructed with the potential interactions.\\
The two Einstein-Hilbert terms include an infinite amount of interactions for $h_{ab}$ and $l_{ab}$
\begin{eqnarray}
\label{EH_term}
-\frac12 M_{\rm Pl}^2 \sqrt{-g}R_g &=& h^{\alpha\beta} \hat{\mathcal{E}}^{\mu\nu}_{\alpha\beta} h\mn + \frac 1 \mpl h (\p h)^2+ \frac 1{\mpl^2} h^2 (\p h)^2 + \cdots \,, \nonumber\\
-\frac12 M_f^2 \sqrt{-f}R_f&=& l^{\alpha\beta} \hat{\mathcal{E}}^{\mu\nu}_{\alpha\beta} l\mn 
+ \frac 1{M_f} l (\p l)^2+ \frac 1{M_f^2} l^2 (\p l)^2 + \cdots \,, 
\end{eqnarray}
whilst the potential only includes a finite number of interactions in $h_{ab}$ and $l_{ab}$
\begin{equation}
\mathcal{U}=-\frac14m^2\mpl^2\sum_{n=0}^4 \frac{\alpha_i}{n!(4-n)!} \mathcal{U}_n[h,l] 
\end{equation}
where the individual potential terms $\mathcal{U}_i$ can be expressed as
\begin{eqnarray}
\label{U2}
\mathcal{U}_0[h] &=& \mathcal E^{abcd}\mathcal E^{a'b'c'd'}(\delta_{aa'}+h_{aa'})(\delta_{bb'}+h_{bb'})(\delta_{cc'}+h_{cc'})(\delta_{dd'}+h_{dd'})\nonumber\\
\mathcal{U}_1[h,l] &=&  \mathcal E^{abcd}\mathcal E^{a'b'c'd'}(\delta_{aa'}+h_{aa'})(\delta_{bb'}+h_{bb'})(\delta_{cc'}+h_{cc'})(\delta_{dd'}+l_{dd'})\nonumber\\
\mathcal{U}_2[h,l] &=&  \mathcal E^{abcd}\mathcal E^{a'b'c'd'}(\delta_{aa'}+h_{aa'})(\delta_{bb'}+h_{bb'})(\delta_{cc'}+l_{cc'})(\delta_{dd'}+l_{dd'}) \nonumber\\
\mathcal{U}_3[h,l] &=&  \mathcal E^{abcd}\mathcal E^{a'b'c'd'}(\delta_{aa'}+h_{aa'})(\delta_{bb'}+l_{bb'})(\delta_{cc'}+l_{cc'})(\delta_{dd'}+l_{dd'}) \nonumber\\
\mathcal{U}_4[l] &=&  \mathcal E^{abcd}\mathcal E^{a'b'c'd'}(\delta_{aa'}+l_{aa'})(\delta_{bb'}+l_{bb'})(\delta_{cc'}+l_{cc'})(\delta_{dd'}+l_{dd'}) 
\end{eqnarray}
where indices are lowered and raised with respect to the flat euclidean metric $\delta_{ab}$. From the five parameters $\alpha_{n}$ we can fix two of them by making the two tadpole contributions for $h_{ab}$ and $l_{ab}$ to vanish:
\begin{equation}
\alpha_1=\frac12(-\alpha_0+2\alpha_3+\alpha_4), \hspace{30pt} \alpha_2=\frac16(\alpha_0-8\alpha_3-3\alpha_4)
\end{equation}
At linear order in perturbations it is trivial to split the mass spectrum of the bigravity theory into the massless and massive spin-2 fluctuations. At the non-linear level and around general backgrounds this split is usually not well-defined \cite{Hassan:2012wr}. However, here we will only need the linear split around euclidean backgrounds in the same way as was done in \cite{deRham:2013qqa}. We will replace in the above potential terms the fluctuations $h_{ab}$ and $l_{ab}$ in terms of the mass modes \ref{fluctuations_massmodes} and compute the leading feynman diagrams one by one and add up their contributions.

\subsection{Tadpole Contributions}
\begin{figure}[!htb]\vspace{20pt}
\begin{center}
\begin{fmffile}{Tadpole_contributions}
$\mathcal{A}^{\rm (1pt)}_{3w}\ \ =$\hspace{10pt}
\parbox{50mm}{\begin{fmfgraph*}(70,30)
	            \fmfleft{i1}
	            \fmfright{o1}
                \fmflabel{$w\mn$}{i1}
                \fmf{wiggly}{i1,v1}
                \fmfdot{v1}
                \fmf{wiggly,left=1,tension=0.8,label=$\psi$}{v1,o1}
                \fmf{wiggly,left=1,tension=0.8}{o1,v1}
\end{fmfgraph*}}\\
$\mathcal{A}^{\rm (1pt)}_{v,2w}\ \ =$\hspace{10pt}
\parbox{50mm}{\begin{fmfgraph*}(70,30)
	            \fmfleft{i1}
	            \fmfright{o1}
                \fmflabel{$v\mn$}{i1}
                \fmf{plain}{i1,v1}
                \fmfdot{v1}
                \fmf{wiggly,left=1,tension=0.8,label=$\psi$}{v1,o1}
                \fmf{wiggly,left=1,tension=0.8}{o1,v1}
\end{fmfgraph*}}\\
$\mathcal{A}^{\rm (1pt)}_{3v}\ \ =$\hspace{10pt}
\parbox{50mm}{\begin{fmfgraph*}(70,30)
	            \fmfleft{i1}
	            \fmfright{o1}
                \fmflabel{$v\mn$}{i1}
                \fmf{plain}{i1,v1}
                \fmfdot{v1}
                \fmf{plain,left=1,tension=0.8,label=$\psi$}{v1,o1}
                \fmf{plain,left=1,tension=0.8}{o1,v1}
\end{fmfgraph*}}\\
$\mathcal{A}^{\rm (1pt)}_{2v,w}\ \ =$\hspace{10pt}
\parbox{50mm}{\begin{fmfgraph*}(70,30)
	            \fmfleft{i1}
	            \fmfright{o1}
                \fmflabel{$v\mn$}{i1}
                \fmf{wiggly}{i1,v1}
                \fmfdot{v1}
                \fmf{plain,left=1,tension=0.8,label=$\psi$}{v1,o1}
                \fmf{plain,left=1,tension=0.8}{o1,v1}
\end{fmfgraph*}}
\end{fmffile}
\end{center}
\caption{Contribution to the graviton tadpole from a graviton loop from the potential interactions. Wiggly lines denote the massless mode of the graviton. The pure tadpole contribution coming from the massless mode $w_{ab}$ is zero in dimensional regularization.}
\label{Feynman_diagrams_Tadpole}
\end{figure}
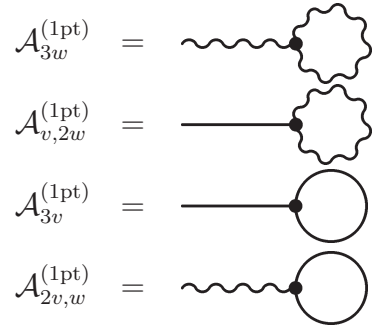

For the tadpole contributions at one loop level, we only have four diagrams to consider as depicted in Figure \ref{Feynman_diagrams_Tadpole}. However, only two of them give a non-trivial contributions. The first two tadpole contributions come from cubic and quadratic interactions in the massless mode $w_{ab}$ which give exactly zero $\mathcal{A}^{\rm (1pt)}_{3w}=0$ and $\mathcal{A}^{\rm (1pt)}_{v,2w}=0$. 
On the other hand, the pure third order interactions in the massive mode 
\begin{equation}\label{pot_3v}
\mathcal{U}_{3v}=-\frac{1}{12}m^2\mpl^2(\alpha_0-\alpha_4) \left( [v]^3 -3[v][v^2]+2[v^3] \right)
\end{equation}
gives rise to the same tadpole contribution for the massive mode as in massive gravity \cite{deRham:2013qqa}
\begin{equation}
\mathcal{A}^{\rm (1pt)}_{3v}=\frac{5}{16}m^4\mpl^2(\alpha_0-\alpha_4)[v]
\end{equation}
 Finally, the contribution coming from the mixed interactions
\begin{equation}
\label{eq:U2v1w}
\mathcal{U}_{2v,w}=-\frac{1}{12}m^2\mpl^2(\alpha_0+4\alpha_3+3\alpha_4)\left( 2(v_a^{\;c}v^{ab}-[v]v^{bc})w_{bc}+(-[v^2]+[v]^2)[w] \right)
\end{equation}
gives a non-trivial new contribution in form of a tadpole for the massless mode
\begin{equation}
\mathcal{A}^{\rm (1pt)}_{2v,w}=\frac{5}{48}m^4\mpl^2(\alpha_0+4\alpha_3+3\alpha_4)[w]
\end{equation}
The one loop contributions coming from cubic order interactions give rise to non-vanishing tadpole contributions for the massive $v_{ab}$ and massless $w_{ab}$ modes. We can now express the massive and massless fluctuations  $v_{ab}$ and $w_{ab}$ back in terms of the fluctuations $h_{ab}$ and $l_{ab}$, which will result in tadpole contributions for $h_{ab}$ and $l_{ab}$.


\subsection{Two-point function Contributions}
\begin{figure}[!htb]
\begin{center}
$\mathcal{A}^{\rm (2pt)}_{2v,w-2v,w}\ \ =$\hspace{10pt}
\begin{fmffile}{Twopoint_3vertex}
\parbox{50mm}{\begin{fmfgraph*}(100,30)
	            \fmfleft{i1}
	            \fmfright{o1}
	           \fmfdot{v1}
                \fmfdot{v2}
                \fmf{plain}{i1,v1}
                \fmf{plain}{v2,o1}
                \fmf{plain,left=0.7,tension=0.4}{v1,v2}
                 \fmf{wiggly,left=0.7,tension=0.4}{v2,v1}
\end{fmfgraph*}}\\
$\mathcal{A}^{\rm (2pt)}_{2v,w-3v}\ \ =$\hspace{1pt}
\parbox{50mm}{\begin{fmfgraph*}(100,30)
	            \fmfleft{i1}
	            \fmfright{o1}
	           \fmfdot{v1}
                \fmfdot{v2}
                \fmf{wiggly}{i1,v1}
                \fmf{plain}{v2,o1}
                \fmf{plain,left=0.7,tension=0.4}{v1,v2}
                 \fmf{plain,left=0.7,tension=0.4}{v2,v1}
\end{fmfgraph*}}\\
$\mathcal{A}^{\rm (2pt)}_{3v-3v}\ \ =$\hspace{10pt}
\parbox{50mm}{\begin{fmfgraph*}(100,30)
	            \fmfleft{i1}
	            \fmfright{o1}
	           \fmfdot{v1}
                \fmfdot{v2}
                \fmf{plain}{i1,v1}
                \fmf{plain}{v2,o1}
                \fmf{plain,left=0.7,tension=0.4}{v1,v2}
                 \fmf{plain,left=0.7,tension=0.4}{v2,v1}
\end{fmfgraph*}}\\
$\mathcal{A}^{\rm (2pt)}_{2v,w-2v,w}\ \ =$\hspace{10pt}
\parbox{50mm}{\begin{fmfgraph*}(100,30)
	            \fmfleft{i1}
	            \fmfright{o1}
	           \fmfdot{v1}
                \fmfdot{v2}
                \fmf{wiggly}{i1,v1}
                \fmf{wiggly}{v2,o1}
                \fmf{plain,left=0.7,tension=0.4}{v1,v2}
                 \fmf{plain,left=0.7,tension=0.4}{v2,v1}
\end{fmfgraph*}}
\end{fmffile}
\end{center}
\caption{1-loop contributions to the 2-point correlation functions with two vertices}
\label{Feynman_diagrams_twopointwv}
\end{figure}
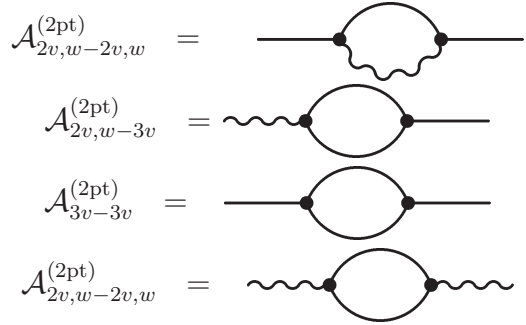

In a similar way we can now compute the two-point functions. There are more diagrams which contribute at the level of the two-point function. Let us start with the Feynman diagrams with 4-vertices giving rise to a 'tadpole 2-point function'. Similarly as before, the diagrams with the massless mode $w_{ab}$ running in the loop will give rise to zero contribution. Thus, the interactions symbolically of the form $\hat{w}^3\hat{v}$ and $\hat{w}^4$ will give zero contributions $\mathcal{A}^{\rm (2pt)}_{4w}=0$ and $\mathcal{A}^{\rm (2pt)}_{v,3w}=0$. \begin{figure}[!htb]
\begin{center}
\begin{fmffile}{Twopoint_tadpole}
$\mathcal{A}^{\rm (2pt)}_{4w}\ \ =$\hspace{10pt}
\parbox{30mm}{\begin{fmfgraph*}(50,50)
		\fmfleft{i1}
		\fmfright{i2}
                  \fmfsurround{i1,i2}
                \fmf{wiggly}{i1,v1}
                \fmf{wiggly}{i2,v1}
                \fmfdot{v1}
                \fmf{wiggly,wiggly=0.5,tension=0.6}{v1,v1}
\end{fmfgraph*}}\\
$\mathcal{A}^{\rm (2pt)}_{v,3w}\ \ =$\hspace{10pt}
\parbox{30mm}{\begin{fmfgraph*}(50,50)
		\fmfleft{i1}
		\fmfright{i2}
                  \fmfsurround{i1,i2}
                \fmf{wiggly}{i1,v1}
                \fmf{plain}{i2,v1}
                \fmfdot{v1}
                \fmf{wiggly,wiggly=0.5,tension=0.6}{v1,v1}
\end{fmfgraph*}}\\
$\mathcal{A}^{\rm (2pt)}_{3v,w}\ \ =$\hspace{10pt}
\parbox{30mm}{\begin{fmfgraph*}(50,50)
		\fmfleft{i1}
		\fmfright{i2}
                  \fmfsurround{i1,i2}
                \fmf{plain}{i1,v1}
                \fmf{wiggly}{i2,v1}
                \fmfdot{v1}
                \fmf{plain,right=0.5,tension=0.6}{v1,v1}
\end{fmfgraph*}}\\
$\mathcal{A}^{\rm (2pt)}_{2v,2w}\ \ =$\hspace{10pt}
\parbox{30mm}{\begin{fmfgraph*}(50,50)
		\fmfleft{i1}
		\fmfright{i2}
                  \fmfsurround{i1,i2}
                \fmf{wiggly}{i1,v1}
                \fmf{wiggly}{i2,v1}
                \fmfdot{v1}
                \fmf{plain,right=0.5,tension=0.6}{v1,v1}
\end{fmfgraph*}} \\
$\mathcal{A}^{\rm (2pt)}_{4v}\ \ =$\hspace{10pt}
\parbox{30mm}{\begin{fmfgraph*}(50,50)
		\fmfleft{i1}
		\fmfright{i2}
                  \fmfsurround{i1,i2}
                \fmf{plain}{i1,v1}
                \fmf{plain}{i2,v1}
                \fmfdot{v1}
                \fmf{plain,right=0.5,tension=0.6}{v1,v1}
\end{fmfgraph*}}
\end{fmffile}
\end{center}
\caption{One-loop contribution to the
2-point correlation function from a graviton internal line coming from quartic interactions.}
\label{Tadpole_2point}
\end{figure}
For the non-trivial contributions, let us first consider the mixed interactions in which there are three $v_{ab}$ modes and one $w_{ab}$ mode coming in
\begin{eqnarray}
\label{eq:U3v1w}
\mathcal{U}_{3v,w}=\frac{1}{12}m^2\mpl^2(\alpha_0-\alpha_4)&&\left( 3(2v_b^{\;d}(v_a^{\;c}v^{ab}-[v]v^{bc})+(-[v^2]+[v]^2)v^{cd})w_{cd} \right. \nonumber\\
&&\left. -(v_{bc}(2v_a^{\;c}v^{ab}-3[v]v^{bc}+[v]^3)[w] \right).
\end{eqnarray}
These interactions will give rise to quantum corrections in the following form \footnote{Note that we also take into account the mirror reflected Feynman diagram by multiplying the result by a factor of two.}
\begin{equation}
\mathcal{A}^{\rm (2pt)}_{3v,w}=\frac{5}{24}m^4\mpl^2(\alpha_0-\alpha_4) \left( [v][w]-v^{ab}w_{ab}\right)
\end{equation}
As next, consider the Feynman diagram with the two massless modes $w_{ab}$ on the external lines and the two massive modes $v_{ab}$ running in the loop with the corresponding interactions given by
\begin{eqnarray}
\label{eq:U2v2w}
\mathcal{U}_{2v,2w}&&=\frac{1}{24}m^2\mpl^2(\alpha_0+4\alpha_3+3\alpha_4)\left(  v^{ab}(2v^{cd}(w_{ac}w_{bd}-w_{ab}w_{cd})+w_{cd}(4v_a^{\;c}w_b^{\;d}-v_{ab}w^{cd}) \right. \nonumber\\
&&\left. +(-4v_a^{\;c}w_{bc}+v_{ab}[w])[w]) +[v](w_{cd}(-4v^{bc}w_b^{\;d}+[v]w^{cd})+(4v^{bc}w_{bc}-[v][w])[w])  \right).
\end{eqnarray}
These interactions give the following non-trivial contribution
\begin{equation}
\mathcal{A}^{\rm (2pt)}_{2v,2w}=-\frac{5}{144}m^4\mpl^2(\beta_0+4\beta_3+3\beta_4) \left( [w^2]-[w]^2 \right)
\end{equation}
The last tadpole two point function is the one corresponding to the massive gravity case, in which namely two massive modes $v_{ab}$ run on the external legs while the other two run in the loop. The interaction is given by
\begin{equation}
\label{eq:U4v}
\mathcal{U}_{4v}=\frac{1}{24}m^2\mpl^2(\alpha_0-4\alpha_3-\alpha_4)\left(  v_{cd}(6v_a^{\;c}v^{ab}v_b^{\;d}-3[v^2]v^{cd}-[v](8v_b^{\;d}v^{bc}-6[v]v^{cd}))-[v]^4 \right).
\end{equation}
The contribution gives the same result as in massive gravity
\begin{equation}
\mathcal{A}^{\rm (2pt)}_{4v}=-\frac{5}{24}m^4\mpl^2(\alpha_0-4\alpha_3-\alpha_4) \left( [v^2]-[v]^2 \right).
\end{equation}
These tadpole diagrams as depicted in Figure \ref{Tadpole_2point} generate quantum corrections which preserve the nice structure of the potential. Their contributions to the counter terms are of the form
\begin{equation}
\mathcal{L}_{CT}=c_1 ([h]^2-[h^2])+c_2([l]^2-[l]^2)+c_3([h][l]-h_{ab}l^{ab})
\end{equation}
where the parameters $c_1,c_2\cdots$etc. are the placeholders for the renormalized parameters $\alpha_n$. Their specific form is irrelevant for now (even though they are important for the purposes of possible exact cancellations for which we took them into account). The important fact is that these one-loop corrections of Figure  \ref{Tadpole_2point} renormalize the potential interactions but do not give rise to detuning. So far, these are excellent news. Actually, exactly the same thing happens in massive gravity (which corresponds to the case where only the last diagram of Figure \ref{Tadpole_2point} contributes) since the tadpole-2 point function does not detune the mass term. However, this optimistic result will not prevail for other corrections, which will indeed detune the specific structure of the potential interactions. To see this, let us now continue with the Feynman diagrams which contain two vertices. They are all shown in Figure \ref{Feynman_diagrams_twopointwv} (we have omitted those diagrams that yield zero contributions). The first diagram constitutes of two vertices with each vertex containing the interaction \ref{eq:U2v1w} and gives the following contribution
\begin{equation}
\mathcal{A}^{\rm (2pt)}_{2v,w-2v,w}=\frac{5}{216}m^4\mpl^4(\alpha_0+4\alpha_3+3\alpha_4)^2\left( 2[v^2]+[v]^2 \right)
\end{equation}
Similarly, the second Feynman diagram contains a vertex with the interaction \ref{eq:U2v1w} while the other vertex being the interaction given in \ref{pot_3v}. Its contribution reads
\begin{equation}
\mathcal{A}^{\rm (2pt)}_{2v,w-3v}=\frac{5}{576}m^4\mpl^4(\alpha_0-\alpha_4)(\alpha_0+4\alpha_3+3\alpha_4)\left( 8[vw]+7[v][w] \right).
\end{equation}

The third diagram on the other hand is constructed purely out of the cubic interaction in $v_{ab}$ \ref{pot_3v}. After performing the integration over the internal momenta, it results in
\begin{equation}
\mathcal{A}^{\rm (2pt)}_{3v-3v}=\frac{5}{32}m^4\mpl^4(\alpha_0-\alpha_4)^2\left( 8[v^2]+7[v]^2 \right).
\end{equation}
Last but not least, the forth diagram in Figure \ref{Feynman_diagrams_twopointwv} generates a contribution of the similar form
\begin{equation}
\mathcal{A}^{\rm (2pt)}_{2v,w-2v,w}=\frac{5}{2592}m^4\mpl^4(\alpha_0+4\alpha_3+3\alpha_4)^2\left( 8[w^2]+7[w]^2 \right).
\end{equation}
As can be seen from the contributions computed above all these two point functions of Figure \ref{Feynman_diagrams_twopointwv} give rise to a detuning of the potential interactions
\begin{equation}
\mathcal{L}_{CT}= (c_1[h]^2-c_2[h^2])+(c_3[l]^2-c_4[l]^2)+(c_5[h][l]-c_6h_{ab}l^{ab})
\end{equation}
where again the parameters $c_1, c_2\cdots$ etc. encode the detuning of the classical parameters at the quadratic order. These parameters are different from the classical parameters. We do not need to compute the contributions to higher n-point functions at this stage, since we already explicitly checked that there is no cancellation happening between the diagrams and exactly the same detuning of the potential interactions happens in massive gravity. These generated quantum corrections to the tadpole and 2-point functions can also not be resumed into cosmological constants. Thus, bigravity seems to share the same destiny as massive gravity and the quantum corrections coming from the graviton loops detune the nice potential interactions and reintroduce the ghost.

\subsection{Scaling of the detuning of the potential interactions}
\label{sec:detuning}
We have explicitly seen above that quantum corrections generated by the graviton loops destroy the very specific structure of the ghost-free potential of massive bigravity exactly in the same way as in massive gravity. The tadpole-2 point functions maintain the potential interactions, however all the other remaining contributions detune the potential and do not cancel with each other nor combine into a cosmological constant. The detuning goes as 
\ba
\label{QC}
\L_{\rm CT}\sim c_i \frac{m^4}{\mpl^i} \ h^i+ d_i \frac{m^4}{M_{\rm f}^i} \ l^i + e_i \frac{m^4}{M_{\rm f}^{i-j}\mpl^j} \ l^{i-j}h^{j}\,,
\ea
We can expand these contributions to quadratic order around the backgrounds $h=\bar h$ and $l=\bar l$, which will detune the Fierz--Pauli structure
\ba
\L_{\rm CT,\ \bar h, \bar l}\sim c_i  \frac{m^4\bar{h}^{i-2}}{\mpl^i}\ h^2 + d_i  \frac{m^4\bar{l}^{i-2}}{M_{\rm f}^i}\ l^2 +  e_i  \frac{m^4\bar{h}^{j-1}}{M_{\rm Pl}^{j}}\  \frac{m^4\bar{l}^{i-j-1}}{M_{\rm f}^{i-j}}\ hl  \, .
\ea
In terms of the helicity-0 degree of freedom this would imply a higher order derivative operator with the mass of the ghost scaling as
\ba
\L_{\rm CT,\ \bar h, \bar l} \sim \frac{ (\p^2 \pi)^2}{m_{\textrm{ghost}}^2}, \hspace{20pt}{\rm with}\hspace{20pt}
m_{\textrm{ghost}}=\(\frac{\mpl}{\bar h}\)^{i/2}\bar h + \(\frac{M_f}{\bar l}\)^{i/2}\bar l +\cdots\,. 
\ea
In the vicinity of small background configurations, the mass of the ghost is very large and hence the ghost is harmless. However, around arbitrarily large background configurations the mass of the ghost can be made arbitrarily small, which a priori is a problem. Nevertheless, this is not the end of the story. For large background configurations, the Vainshtein mechanism needs to be inserted at the quantum level. We expect that the Vainshtein mechanism will repackage the one-loop effective action and suppress the quantum corrections around large background configurations, exactly in the same way as in massive gravity. A detail investigation of this is out of the scope of this work. The mathematically challenging computation of the one-loop effective action with the Vainshtein mechanism implemented will be studied somewhere else.


\section{Matter loops}
\label{sec:MatterLoops}
In massive (bi-)gravity the existence of the two metrics comes hand in hand with the natural question of how these two metrics can couple to the matter sector consistently. First this has to be established  successfully at the classical level and then as a following step one needs to make sure that this property can be further extended to the quantum level. However, this will not be the philosophy that we will be following here. We will follow the same logic as in \cite{deRham:2014naa} and demand the requirement of quantum stability to deduce the possible ways of coupling to matter fields. If the matter field couples to only one metric the classical theory is free of any ghost instability \cite{Hassan:2011zd}. This property is also maintained at the quantum level since the quantum corrections do not renormalize the potential interactions or detune them but rather contribute in form of a cosmological constant \cite{deRham:2014naa}. This will be one of the valid couplings that we will be considering here. We will disregard the case in which the matter field couples to both metrics at the same time, since for this coupling there is a ghost degree of freedom already present at the classical level \cite{deRham:2014naa} and the quantum corrections do detune the potential interactions. We will consider the case in which new effective composite metrics can be constructed from lessons learned from quantum corrections and to which the matter field can couple to both metrics simultaneously.

\subsection{Coupling to separate matter sector}
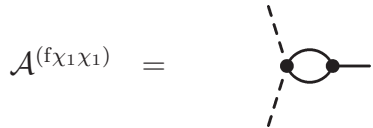
\begin{figure}[!htb]\vspace{10pt}
\begin{center}
$\mathcal{A}^{\rm (f\chi_1\chi_1)}\ \ =$\hspace{15pt}
\begin{fmffile}{Scattering_Coupling}
\parbox{20mm}{\begin{fmfgraph*}(50,50)
	            \fmfsurround{i1,i2,i3}
                \fmf{plain}{i1,v1}
                \fmf{dashes}{i2,v2}
                \fmf{dashes}{i3,v2}
                \fmfdot{v1,v2}
                \fmf{plain,left=0.7,tension=0.4}{v1,v2,v1}
\end{fmfgraph*}}
\end{fmffile}
\end{center}
\caption{One-loop contributions to the 3-point function $f_{ab}\delta^{ab}\chi_1^2$. Dashed lines denote the matter field $\chi_1$.}
\label{Feynman_diagrams_couplins}
\end{figure}

In massive gravity the dynamical metric $g_{ab}$ can be coupled covariantly to the matter sector without altering the number of propagating degrees of freedom. This nice property remains valid at the quantum level as well since the quantum corrections give rise to a contribution in form of a cosmological constant \cite{deRham:2013qqa}. Therefore a promising way of coupling the two metrics in bigravity is through an independent coupling to separate matter sector  
\ba
\L_{\rm matter}= \frac 12 \sqrt{g} \(g^{ab}\p_a \chi_1 \p_b \chi_1 +M_1^2 \chi_1^2\)
+\frac 12 \sqrt{f} \(f^{ab}\p_a \chi_2 \p_b \chi_2 +M_2^2 \chi_2^2\)\,.
\ea
where we assumed massive scalar fields as matter fields for simplicity. The two scalar fields $\chi_1$ and $\chi_2$ with masses $M_1$ and $M_2$ couple separately to $g_{ab}$ and $f_{ab}$ respectively, but not to both simultaneously.
Similary to what happens in massive gravity the contributions to the one loop effective action are in form of additive cosmological constants for $g_{ab}$ and $f_{ab}$ \cite{deRham:2014naa}
\ba
\mathcal{L}^{({\rm matter-loops})}_{1, \log }=M_1^4 \sqrt{g}\log(M_1/\mu) +M_2^4\sqrt{f} \log(M_2/\mu) + \text{curvature corrections}\,,
\ea
Even if we force at the classical level that only the $g_{ab}$ metric couples to the matter field $\chi_1$, i.e. there is no coupling between the metric $f_{ab}$ and the matter field $\chi_1$, it is an unavoidable question we have to pursue whether or not quantum corrections will generate couplings between $f_{ab}$ and $\chi_1$ and if so at which scale they become important. Diagrams as shown in figure \ref{Feynman_diagrams_couplins} will indeed generate new coupling between $f_{ab}$ and the matter field $\chi_1$
   \\
At the one vertex the interactions coming from the potential at quadratic order in $h_{ab}$ and linear order in $l_{ab}$ will contribute. We shall keep in mind that the potential interactions have a very specific antisymmetric structure and the one acting in the above diagram has the form $\E^{abcd}\E^{a'b'c'}_{\ \ \ \ \  d} h_{a a'}h_{bb'}l_{cc'}$. And at the other vertex the coupling to matter field expanded to quadratic order in $h$ will contribute as
\begin{equation}\label{matter_gcoupling}
\left(\frac32h^a_c h^{c b}-[h]h^{ab}+\frac14([h]^2-[h^2])\delta^{ab} \right)\partial_a\chi_1 \partial_b\chi_1+\frac{M^2\chi_1^2}4([h]^2-[h^2])
\end{equation}

We can contract the $l_{ab}$ field of the potential interaction with an external $l_{ab}$ leg coming out of this vertex while the other two $h_{ab}$-fields from this vertex run in the loop with momenta $k_a$ and $(p_1+p_2-k)_a$ and contract them with the two spin-2 fields coming from the coupling \ref{matter_gcoupling}. Strictly speaking it is not the $h_{ab}$ field which is running in the loop but the massless and massive modes which are diagonal. So the propagator for the $h_{ab}$ field $\langle h_{ab}h_{cd}\rangle$ needs to be replaced by the some of the propagators of the massless mode $w_{ab}$ and massive mode $v_{ab}$. Since these two modes are diagonal there will not be any mixing of the form $\langle w_{ab}v_{cd}\rangle$. One could worry that since the gravitons are running in the loops that the contribution of this diagram might scale with an inverse power of $m$ coming from the propagator of the massive mode $\langle v_{ab}v_{cd}\rangle$. Since we have two graviton propagators in this diagram, if each internal propagator comes at least with $k^4/m^4$ from the massive mode than the contribution of this vertex to the graph with the most negative powers of the graviton mass would be $m^{-8}$
\begin{eqnarray}
\mathcal A&&
 \propto  \int \frac{\d^4 k}{(2\pi)^4} \E^{abcd}\E^{a'b'c'}_{\ \ \ \ \ d} \frac{k_ak_{a'}}{m^2}\frac{k^fk_e}{m^2}\frac{(p_1+p_2-k)_b (p_1+p_2-k)_{b'}}{m^2} \nonumber \\
&& \times \frac{(p_1+p_2-k)^e  (p_1+p_2-k)^g}{m^2}\left(p_{1f}p_{2g}\right) l_{cc'}\cdots
\end{eqnarray}
Due to the antisymmetric structure of the interactions $\E^{abcd}\E^{a'b'c'}_{\ \ \ \ \  d}$, this contribution cancels exactly. The Levi-Civita tensors are antisymmetric while the momenta are symmetric.  Exactly the same happens to the contribution with the $m^{-6}$ scaling. Nevertheless, this argument applies only to the contributions with $m^{-6}$ and $m^{-8}$ scalings since there are enough momenta which give rise to symmetric contributions. Actually, this will be true for any n-point function. We will have diagrams which contribute to the n-point function with the highest number of internal graviton propagators n+1 scaling with the most negative power of $m^{-4(n+1)}$. However, they will cancel exactly as in the above diagram. The cubic vertex from the potential term gives zero to leading order and $m^2$ scaling to second leading order. Therefore for the n-point function, each vertex cannot contribute with more than $m^{-2}$, meaning that the divergence  is at worst like $m^{-2(n+1)} $rather than $m^{-4(n+1)}$. Exactly the same reasoning applies to quartic and higher dimensional vertices in the $h$ field. Each internal propagator comes at least with $k^2/m^2$. In the case in which each propagator contributes with $k^2/m^2$ would give rise to  $(k/m)^{2n}$ contributions in the vertex which is fully symmetric and cancel do to the antisymmetric structure of the potential term with the Levi-Civita tensors.\\

Returning to our diagram, we explicitly saw that the contributions scaling with $m^{-6}$ and $m^{-8}$ powers cancel. However, the above argumentation does not apply on the contributions with $m^{-2}$ and $m^{-4}$ scalings, and we need to check their implications. Unfortunately, there are indeed contributions with such a scaling. Let us investigate how the dependence on the $m$ scaling will change the scale at which these interactions between $f_{ab}$ and the matter field $\chi_1$ will become important. The above diagram will give rise to interactions between $f_{ab}$ and the matter field $\chi_1$ at the scale
\begin{equation}\label{matter_induced_fcoupling}
m^2 M_{\rm Pl}^2\frac{1}{M_f}\frac{1}{M_{\rm Pl}^2} \frac{1}{M^2_{\rm Pl}} \frac{k^{2(n+1)}}{m^{2n}} f_{ab}\delta^{ab} \chi_1 \chi_1
\end{equation}
Now, we can compute the corrections coming from these new interactions (which we know that they will give rise to a ghost degree of freedom). Consider a diagram in which $h_{ab}$ and $l_{ab}$ run on the external legs while the $\chi_1$ field runs in the loop. Such diagrams will scale as
\begin{equation}
\frac{M_1^2}{M_{\rm Pl}}m^{2(1-n)}M_{\rm Pl}^2\frac{1}{M_f}\frac{1}{M_{\rm Pl}^2}\frac{1}{M^2_{\rm Pl}} M_1^{2(n+1)}
\end{equation}
Without loss of generality, assume for clarity that $M_f=M_{\rm Pl}$, then the ghost associated with the higher derivative operators applied on the Stueckelberg field would come at a scale
\begin{equation}
\frac{M_1^2}{M_{\rm Pl}}m^{2(1-n)}\frac{1}{M_{\rm Pl}^3}\frac{M_1^{2(n+1)}}{m^4}
\end{equation}
meaning that the mass of the ghost would correspond to
\begin{equation}
m^2_{\rm ghost}=m^{-2(1+n)}M_{\rm Pl}^4\frac{1}{M_1^{2(2+n)}}
\end{equation}
We have seen above explicitly that the contributions with $n\ge 3$ cancel exactly due to the antisymmetric structure of the potential interactions. The only two cases we need to check are $n=1$ and $n=2$. Let us assume, that the mass of the matter field is close to $M_1 \approx \Lambda_3$. Then, for $n=1$ we would have $m^2_{\rm ghost}=\frac{\Lambda_3^6M^{2}_{\rm Pl}}{\Lambda_3^6}$ which is larger than the strong coupling scale $\Lambda_3$. Similarly, for $n=2$ we would obtain $m^2_{\rm ghost}=\frac{\Lambda_3^9M_{\rm Pl}}{\Lambda^8_3}$ which is as well beyond the scale $\Lambda_3$. Thus, even if the quantum corrections reintroduce a coupling between $f_{ab}$ and $\chi_1$ which was put to zero at the classical level, the scaling of this new coupling would yield a ghost well beyond the strong coupling scale.

\subsection{Coupling the matter sector to both metrics}
If one insists on coupling the matter sector to the two metrics $g_{ab}$ and $f_{ab}$ at the same time, the quantum corrections restrict the possible ways crucially. If the quantum corrections detune the very specific potential structure, then the ghost degree of freedom reappears with a scaling that can be made arbitrarily small by choosing the mass of the matter field accordingly. 
In \cite{deRham:2014naa} a new type of coupling to matter was proposed. The coupling occurred through an effective composite metric $g_{\rm eff}$ built out of both metrics $g_{ab}$ and $f_{ab}$ 
\begin{eqnarray}
\label{eq:geffintro}
g^{\rm eff}_{ab}=\alpha^2 g_{ab} +2 \alpha \beta\ g_{a c} (\sqrt{g^{-1}f})^c_{\ b} +\beta^2 f_{ab}\,,
\end{eqnarray}
with arbitrary constants $\alpha$ and $\beta$. The one-loop contributions through matter loops do not contribute in form of a cosmological constant with respect to $g_{ab}$ or $f_{ab}$ but rather with respect to this composite metric $\sqrt{\det{g^{\rm eff}_{ab}}}$ and it was constructed by the requirement that it corresponds to the ghost-free potential interactions of massive (bi-)gravity. In the following we will propose yet other class of new effective composite metrics to which the matter fields can couple and not reintroduce the ghost-freedom at the quantum level. 
\subsubsection{Contributions in form of cosmological constants}
One possible way of constructing the effective composite metric, which was not considered in \cite{deRham:2014naa}, comes from the additive contributions of the cosmological constants for $g_{ab}$ and $f_{ab}$. One has to demand that the determinant of the effective metric is such that it fulfills the following relation
\begin{equation}\label{additionalDet}
\sqrt{\det{\hat{g}_{\rm eff}}}=\sqrt{\det \hat{g}}+\sqrt{\det \hat{f}}
\end{equation} 
If the matter sector would couple to an effective composite metric with a determinant as given in \ref{additionalDet}, then the quantum corrections would not render the theory unnatural. The contributions of matter loops would be the sum of the cosmological constants for $g_{ab}$ and $f_{ab}$ and would not renormalize the potential interactions. The naturalness of massive gravity is one of the essential strength of massive gravity and it would unfortunate to loose this nice property. The above relation \ref{additionalDet} allows only for those type of composite effective metrics that maintain the naturalness property of massive gravity. The solution for the effective composite metric will be then simply given by
\begin{equation}\label{additiveGeffsol}
\hat{g}_{\rm eff}=\left( \sqrt{\det \hat{g}}+\sqrt{\det \hat{f}} \right)^{1/2}\hat{M}
\end{equation}
with an arbitrary matrix $\hat{M}=M_{ab}$ with its determinant fixed to be one, $\det(M_{ab})=1$. This requirement only fixes one of the components of the matrix $\hat{M}$ such that one has a nine parametric solution to the above equation \ref{additionalDet}. The most trivial solution would be $\hat{M}=\delta_{ab}$, in this case the effective composite metric would be simply given by \footnote{Note that we are working in the Euclidean space in this work. If one switches to the Lorenzian space then the effective metric needs to be changed accordingly to the Lorenzian signature \\ $g^{\rm eff}_{\mu\nu}=\left(\sqrt{-\det \hat{g}}+\sqrt{-\det \hat{f}}  \right)^{1/2}\eta_{\mu\nu}$. }
\begin{equation}
g^{\rm eff}_{ab}=\left(\sqrt{\det \hat{g}}+\sqrt{\det \hat{f}}  \right)^{1/2}\delta_{ab}
\end{equation}
Other perfectly valid solutions would be for instance 
\begin{eqnarray}
\hat{M}=\frac{\hat{g}}{ \left(\sqrt{\det \hat{g}}\right)^{1/2}}, \qquad  \hat{M}=\frac{\hat{f}}{ \left( \sqrt{\det \hat{f}}\right)^{1/2}},
 \qquad  \hat{M}=\frac{\gamma_1\hat{g}+\gamma_2\hat{f}+\gamma_3\hat{g}\sqrt{\hat{g}^{-1}\hat{f}}}{\left( \det\left(  \gamma_1 \hat{g}+\gamma_2 \hat{f}+\gamma_3 \hat{g}\sqrt{\hat{g}^{-1}\hat{f}} \right)\right)^{1/4}}  \nonumber
 \end{eqnarray}
for which the new effective metric would then correspond to
\begin{eqnarray}
g^{\rm eff}_{ab}&=&\left(\sqrt{\det \hat{g}}+\sqrt{\det \hat{f}}  \right)^{1/2} \frac{g_{ab}}{ ( \sqrt{\det \hat{g}})^{1/2}} \nonumber \\
g^{\rm eff}_{ab}&=&\left(\sqrt{\det \hat{g}}+\sqrt{\det \hat{f}}  \right)^{1/2} \frac{f_{ab}}{ \left( \sqrt{\det \hat{f}}\right)^{1/2}}   \nonumber \\
 g^{\rm eff}_{ab}&=&\left(\sqrt{\det \hat{g}}+\sqrt{\det \hat{f}}   \right)^{1/2}\frac{\gamma_1\hat{g}+\gamma_2\hat{f}+\gamma_3\hat{g}\sqrt{\hat{g}^{-1}\hat{f}}}{\left( \det\left(  \gamma_1 \hat{g}+\gamma_2 \hat{f}+\gamma_3 \hat{g}\sqrt{\hat{g}^{-1}\hat{f}} \right)\right)^{1/4}} 
\end{eqnarray}
These are only some examples we mention. We can construct any arbitrary tensor of the form 
\begin{equation}
\hat{M}=\frac{\hat{N}}{\det(\hat{N})^{1/4}}
\end{equation}
for which $\det(\hat{M})=1$ would be guaranteed. The tensor $\hat{N}$ could be any combination of the form $\hat{N}=\gamma_1\hat{g}+\gamma_2\hat{f}+\gamma_3\hat{g}\sqrt{\hat{g}^{-1}\hat{f}}+\gamma_4\hat{f}\sqrt{\hat{g}^{-1}\hat{f}}+\gamma_5\hat{g}\sqrt{\hat{f}^{-1}\hat{g}}+\gamma_6\hat{f}\sqrt{\hat{f}^{-1}\hat{g}}\cdots$etc, constructed out of $\hat{g}$ and $\hat{f}$. Thus, any effective metric of the form \ref{additiveGeffsol} with an arbitrary tensor  $\hat{M}$ with $\det(\hat{M})=1$ will be a valid solution. All these effective composite metrics give rise to quantum contributions in form of cosmological constants for $f_{ab}$ and $g_{ab}$ and thus fulfill our requirement in \ref{additionalDet} and do not destroy the naturalness of the theory. The above construction guaranties the quantum stability under matter loops. However, even if the quantum corrections do not introduce any ghost instability, it does not mean that the theory is free from the BD ghost at the classical level. Among all these possible effective metrics most of them will probably excite the BD ghost. All these new effective metrics need to be carefully studied at the classical level, which is out of scope of this work and will be studied somewhere else. Even if these couplings turn out to be not completely free of the BD ghost, it would be interesting to study whether or not the decoupling limit is free of the BD ghost, and what the mass of the ghost is exactly, such that the theory could be considered as an effective field theory.

\subsubsection{Contributions in form of potential interactions}
The other possible way of constructing the effective metric, which was the criteria used in \cite{deRham:2014naa}, corresponds to demanding that the quantum corrections of matter loops are in form of the allowed ghost-free potential interactions
\begin{equation}\label{potentialDetgeff}
\sqrt{\det{\hat{g}_{\rm eff}}}=\sqrt{\det \hat{g}}\det(\alpha \mathbbm 1+\beta \hat{X})
\end{equation}
where $\hat{X}$ stands for $\hat{X}=\sqrt{\hat{g}^{-1}\hat{f}}$. In this way the quantum corrections would not detune the potential interactions and hence introduce ghost degrees of freedom, however they would renormalize the potential interactions and one would loose the naturalness argument. Again we can find the solutions for the effective metric which fulfills the relation \ref{potentialDetgeff}. The generic solution will be of course simply of the form
\begin{equation}\label{solgeffgen}
\hat{g}_{\rm eff}=  \hat{g} (\alpha+\beta \hat{X})^2 \mathcal{\hat{M}}
\end{equation}
with again an arbitrary matrix $\mathcal{\hat{M}}=\mathcal{M}^a_{\; b}$ with the determinant $\det(\mathcal{M}^a_{\; b})=1$. The simplest case with $\hat{\mathcal{M}}= \mathbbm 1$ was the one that was considered in \cite{deRham:2014naa} and gave rise to the effective metric in \ref{eq:geffintro}. Any solution of the form \ref{solgeffgen} with $\det(\hat{\mathcal{M}})=1$ would fulfill the relation \ref{potentialDetgeff}. Thus, we again have a nine parametric solution. For any arbitrary matrix $\hat{\mathcal{M}}$ with $\det(\hat{\mathcal{M}})=1$ the effective metric would be given by \ref{solgeffgen} where among all these solutions the simplest would be given as in the above section
\begin{equation}
\hat{\mathcal{M}}= \mathbbm 1,\qquad \hat{\mathcal{M}}=\frac{\sqrt{\hat{g}^{-1}f}}{ \det\left(\sqrt{\hat{g}^{-1}f} \right)^{1/4}}, \qquad  \hat{\mathcal{M}}=\frac{\sqrt{\hat{f}^{-1}g}}{ \det\left(\sqrt{\hat{f}^{-1}g} \right)^{1/4}}
\end{equation}
Let us emphasize again that in general it can be any matrix fulfilling $\hat{\mathcal{M}}=\frac{\hat{N}}{\det(\hat{N})^{1/4}}$. \\
These new effective composite metrics to which the matter field can couple do not introduce the BD ghost at the quantum level. However, as mentioned above, again one has to study carefully whether or not there is one (or several) of them which is also ghost-free at the classical level. Even in the presence of ghost degrees of freedom, it would be crucial to study the exact mass of the ghost and whether or not they can be considered as an effective field theory with the cutoff scale given by the mass of the ghost. These constitute new avenues to explore that we propose here and it might offer new interesting phenemenology, which shall be studied in great detail in future works.

\section{Conclusions}
\label{sec:conclusion}
This work  was dedicated to the study of quantum corrections in massive bigravity. Starting with the leading interactions in the decoupling limit, we could generalize the non-renormalization theorem to the case of bigravity. The decoupling limit of bigravity is safe from quantum corrections. Beyond the decoupling limit, if we consider only one loop contributions coming from the interactions with the matter fields, they will only yield a contribution in terms of a cosmological constant exactly as in massive gravity if the two metrics are coupled to different matter fields. In case the matter fields couple to both metrics at the same time, then the destabilization of the potential is unavoidable and the mass of the matter fields could be chosen such that the associated ghost appear below the strong coupling scale $\Lambda_3$  \cite{deRham:2014naa} . The same is true if the matter fields couple to different metrics but they interact with each other. Knowing the exact behavior of quantum corrections through matter loops, one can construct an effective composite metric through which the matter sector can couple to both metrics. Following the lessons learned in \cite{deRham:2014naa} we proposed yet other types of effective metrics which either give rise to contributions in form of the cosmological constants for the two metrics (and hence maintaining the naturalness of the theory) or in form of the allowed ghost-free potential interactions. These new composite metrics could give rise to consistent theories at the classical level, which should be carefully studied in future works. Furthermore, we have studied the quantum corrections coming from purely graviton loops. Since we have two dynamical metrics, there will be one massless and one massive spin-2 field running in the loops. We were able to show that the structure of the interactions between the two metrics gets destabilized through graviton loops, exactly in the same way as in massive gravity. It would be an interesting question to pursue whether or not the mass of the ghost can be pushed below the strong coupling scale around arbitrarily large backgrounds. For that purpose, one has to compute the one loop effective action with the Vainshtein mechanism implemented in it. We expect a similar behavior as in massive gravity. 

\acknowledgments

We would like to thank Jose Beltran Jimenez, Claudia de Rham, A. Emir G\"umr\"uk\c c\"uo\u glu, Fawad Hassan, Andrew Matas and Shinji Mukohyama for very useful and enlightening discussions. 
We acknowledge the use of the xAct package for
Mathematica \cite{Brizuela:2008ra, xAct}.


	\bibliographystyle{JHEPmodplain}
	\bibliography{references_QC_GravitonLoops}

\end{document}